\documentclass[twocolumn,superscriptaddress,nobibnotes,nofootinbib]{revtex4-2} 
\usepackage[utf8]{inputenc}
\usepackage{amsmath}
\usepackage[english]{babel}
\usepackage[dvipsnames]{xcolor}
\usepackage{amssymb}
\usepackage{textcomp}
\usepackage{graphicx}
\usepackage{placeins}
\usepackage{textgreek}
\usepackage{upgreek}
\usepackage{latexsym}
\usepackage{braket}
\usepackage[nolist]{acronym}
\usepackage{siunitx}
\usepackage{xpatch}
\usepackage{dsfont}
\usepackage{hyperref}
\usepackage[dvipsnames]{xcolor}
\usepackage{bbding,enumitem}
\usepackage{amsmath}

\newcommand{\M}{M\"ossbauer }

\hypersetup{linktoc=all, hyperindex=true,colorlinks, linkcolor={red!50!black},citecolor={blue!50!black},urlcolor={blue!80!black}}

\makeatletter

\def\l@subsubsection#1#2{}
\patchcmd{\@ssect@ltx}
    {\addcontentsline{toc}{#1}{\protect\numberline{}#8}}
    {}
    {}
    {}
\makeatother

\usepackage{xcolor}
\usepackage{soul}
\sethlcolor{yellow}
\soulregister\ref7
\soulregister\pageref7
\soulregister\eqref7
\soulregister\cite7
\soulregister\M7
\usepackage[most]{tcolorbox}
\tcbuselibrary{breakable}

\newtcolorbox{hlbox}{
  enhanced,
  breakable,
  parbox=false,        
  colback=yellow,
  colframe=yellow,
  boxrule=0pt,
  arc=0pt,
  outer arc=0pt,
  left=0pt,
  right=0pt,
  top=-10pt,
  bottom=0pt,
  before skip=0pt,
  after skip=0pt,
}

\begin{document}

\title{Coupling of a Nuclear Transition to a Surface Acoustic Wave}

\author{Albert Nazeeri}
\affiliation{Department of Physics, Stanford University, Stanford, California 94305, USA}

\author{Chiara Brandenstein}
\affiliation{Department of Physics, Stanford University, Stanford, California 94305, USA}

\author{Chengjie Jia}
\affiliation{Department of Physics, Stanford University, Stanford, California 94305, USA}

\author{Lorenzo Magrini}
\email{magrini@stanford.edu}
\affiliation{Department of Physics, Stanford University, Stanford, California 94305, USA}

\author{Giorgio Gratta}
\affiliation{Department of Physics, Stanford University, Stanford, California 94305, USA}
\affiliation{Hansen Experimental Physics Lab, Stanford University, Stanford, California 94305, USA}

\begin{abstract}
Mechanical modulation of recoilless nuclear transitions allows the dynamic control of $\gamma$-ray emission and absorption. Accessing modulation frequencies well above the nuclear linewidth enables coherent manipulation of the nuclear response. Here we demonstrate high frequency control via efficient coupling a film of enriched $^{57}$Fe to a $97.9~\mathrm{MHz}$ surface acoustic wave, nearly two orders of magnitude higher than the nuclear linewidth. The mechanical drive produces a comb of absorption sidebands in the \M spectrum, reflecting the periodic time modulation of the nuclear transitions. This constitutes the highest frequency mechanically driven \M resonance to date. Our solid-state, monolithic platform establishes a new interface between nuclear transitions and high-frequency acoustics, with applications in $\gamma$-ray quantum optics and precision nuclear spectroscopy.
    
\end{abstract}

\maketitle

\section*{Introduction}
When low-energy nuclear $\gamma$ transitions occur in certain solids, the recoil can be coherently absorbed by the entire lattice. This results in a recoil-free transition that is characterized by extremely narrow spectral lines and is known as the \M effect~\cite{mossbauer_kernresonanzabsorption_1958}. Discovered in 1958, the \M effect was soon employed in a landmark experiment to measure the gravitational redshift, one of the predictions of general relativity~\cite{pound_apparent_1960}.
Today, \M spectroscopy is a useful tool in chemistry and materials science; shifts of the extremely narrow resonances are sensitive local probes to the electric, magnetic, and structural environments of \M active atoms.~\cite{gutlich_mossbauer_2011}.
Beyond precision spectroscopy, the \M effect is a fundamental tool in nuclear and x-ray quantum optics~\cite{adams_x-ray_2013,rohlsberger_electromagnetically_2012,haber_collective_2016,lentrodt_toward_2025}. Coherent control of a nuclear ensemble, achieved through external magnetic fields~\cite{heiman_rf-induced_1968} or mechanical modulation~\cite{ruby_acoustically_1960,mketchyan_modulation_1979,yamashita_measurement_2024}, has enabled demonstrations such as waveform shaping of recoilless $\gamma$-ray photons~\cite{vagizov_coherent_2014} and spectral narrowing of synchrotron pulses~\cite{heeg_spectral_2017}.
Here, we demonstrate the coupling of enriched ${}^{57}\mathrm{Fe}$ nuclei to a traveling surface acoustic wave (SAW) at $97.9~\mathrm{MHz}$ in a monolithic device.  This approach offers three fundamental advantages over bulk piezo driven films. First, it enables modulation of nuclear resonances at frequencies well above the nuclear linewidth, extending the accessible modulation bandwidth beyond what has been achieved with conventional piezo-driven schemes~\cite{mketchyan_modulation_1979,vagizov_coherent_2014,heeg_spectral_2017,yamashita_measurement_2024}. 
Second, the fabrication of surface acoustic wave devices is well-controlled and scalable, with considerable flexibility in the design. Third, SAWs constitute a mature solid-state platform well understood in the quantum regime ~\cite{schuetz_universal_2015,manenti_surface_2016,satzinger_quantum_2018} and compatible with a wide variety of quantum systems~\cite{sletten_resolving_2019,golter_optomechanical_2016,patel_surface_2024}.

We observe Floquet sidebands of the nuclear transitions, with intensities following the characteristic Bessel-function dependence on the modulation strength. To our knowledge, this represents the highest-frequency phonon-driven modulation of \M resonances reported to date, and the first realized in a monolithic device, establishing a new method for high-bandwidth, time-resolved mechanical control of nuclear resonances.

\section*{Experiment}
\begin{figure}[]
    \centering
    \includegraphics[width=8.6cm]{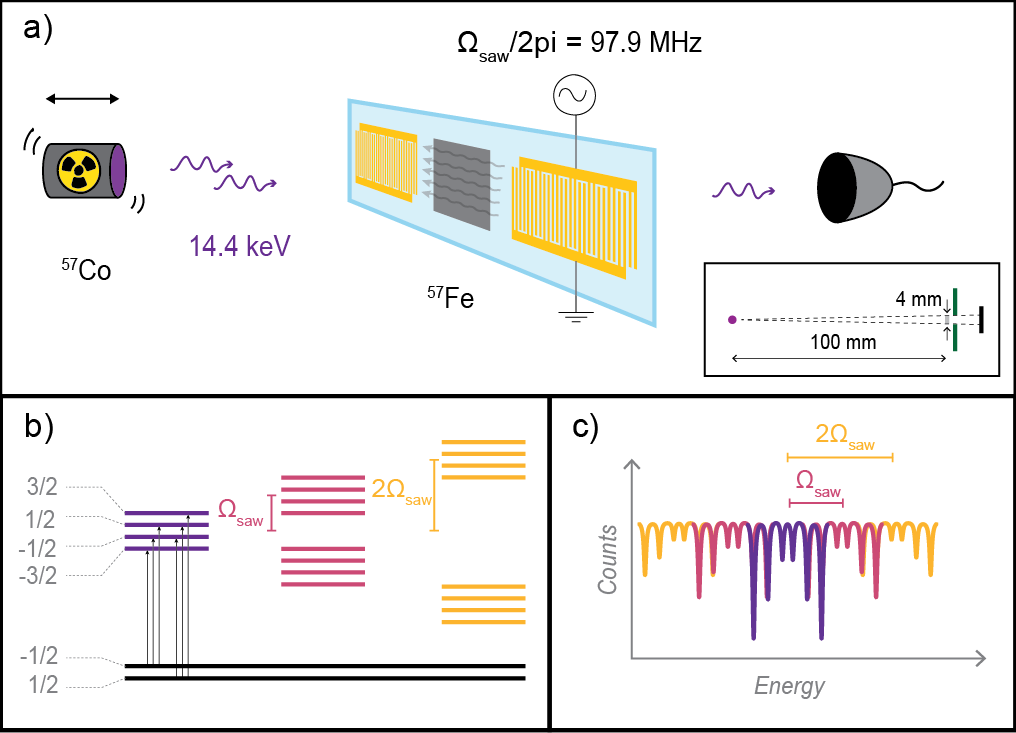}
    \caption{a) Schematic view of the experiment. A $(0.9 \times 4.0)~\mathrm{mm}^2$ film of $^{57}$Fe  is deposited between two SAW couplers on a quartz substrate and characterized in a \M spectrometer. The inset shows the setup to scale, with the SAW device and $^{57}$Fe absorbing film (gray) placed at a distance of $10~\mathrm{cm}$ from the $^{57}$Co source (purple). A tungsten collimator (green) blocks all non-interacting photons from reaching the detector (black). b) Nuclear level scheme, showing the the unmodulated hyperfine transitions (purple) and the first (magenta) and second (yellow) order vibrational sidebands driven by the SAW. c) Cartoon of the expected spectrum without (purple) and with (all colors) the SAW drive on.}
    \label{fig:figure1}
\end{figure}

To enable high frequency mechanical modulation of nuclear resonances, an $^{57}\mathrm{Fe}$ film  is deposited on a ST-Cut quartz substrate, and the film is driven by a $97.9~\mathrm{MHz}$ surface acoustic wave. This modulated \M absorber is placed in a conventional transmission spectrometer~\cite{spectrometer} (Figure~\ref{fig:figure1}a).
The \M absorber is fabricated by thermally evaporating iron enriched to 96\% in the isotope $^{57}\mathrm{Fe}$, forming a $200~\mathrm{nm}$-thick film. The deposition is obtained in a dedicated setup at a pressure of $9.5 \cdot10^{-10}~\mathrm{Torr}$ and a deposition rate of $6.1~\mathrm{nm/h}$.
This fully monolithic implementation, using surface acoustic waves to avoid intermediate, lossy adhesive layers, concentrates the mechanical energy near the surface, resulting in larger Doppler velocities per unit power compared to bulk acoustic transducers.

\begin{figure}[h!]
    \centering
    \includegraphics[width=8.6cm]{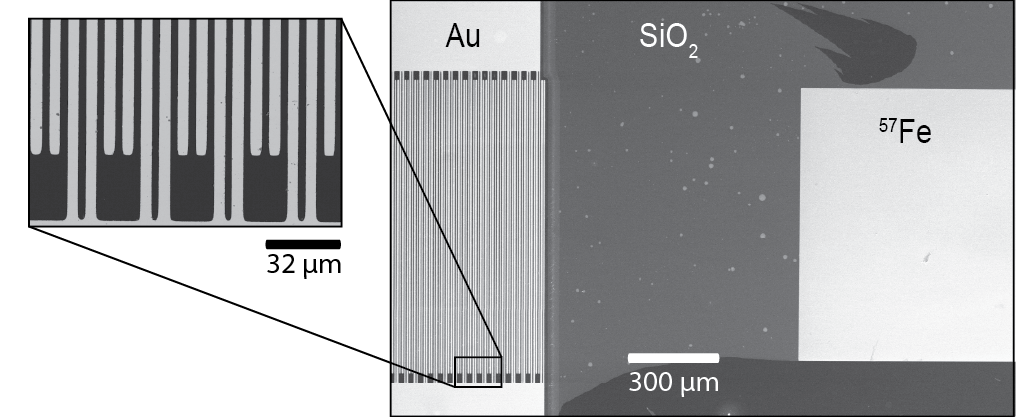}
    \caption{Main panel: Microphotograph of part of the absorber device on the quartz substrate, including sections of the $^{57}$Fe film and of one of the SAW couplers. Inset: higher magnification detail of one of the SAW couplers.}
    \label{fig:figure2}
\end{figure}

With no SAW excitation, the absorption spectrum is dominated by hyperfine Zeeman interactions in the ${}^{57}\mathrm{Fe}$ film which splits the nuclear ground (excited) state into 2 (4) hyperfine states of angular momentum $m_j = \pm1/2$ ($m_j =\pm1/2, \pm3/2$) (purple lines in Fig.\ref{fig:figure1}b-\ref{fig:figure1}c). The selection rule $\Delta m_j = 0,\pm1$ then leads to 6 different allowed transitions~\cite{gutlich_mossbauer_2011}, spaced by 101~neV (corresponding to $24.4~\mathrm{MHz}$, or a Doppler velocity of $2.1~\mathrm{mm/s}$), schematically shown in the bottom panel of Figure~\ref{fig:figure1}b. The data shown in this letter, was taken on a device where over etching in the final step resulted into a reduction of the total amount of iron, so that the effective thickness was approximately $90$~nm over a region of $(0.9 \times 4.0)~\mathrm{mm}^2$ (thickness estimated by the contrast of the peaks in the absorption spectrum)~\cite{appendix}.
SAWs are excited in the quartz by two sets of interdigitated transducers (IDTs or ``couplers'') deposited at opposite ends of the chip, with a pattern of double lines, spaced by a $\lambda_\mathrm{saw} = 32~\mathrm{\mu m}$ pitch, illustrated in Figure~\ref{fig:figure2}. The pattern is designed to resonantly couple, by piezoelectric effect, $\Omega_\mathrm{saw} = 2\pi\cdot 97.9~\mathrm{MHz}$ SAWs propagating across the chip. The two opposite couplers allow for testing the coupling from electric signals, to acoustics, and back to electric signals. To avoid reflections off the edge of the substrate, the edges of the quartz chip are ``terminated'' by small pieces of kapton tape. A fraction, $\alpha$, of the SAW is still coherently reflected by the impedance mismatch from the readout coupler. This reflection gives rise to a partial standing wave across the quartz chip. The resulting modulation of the SAW amplitude is included in the spectral analysis.
A $1$~GBq $^{57}$Co source was used for the experiment, where the $^{57}$Co is embedded in a rhodium matrix, resulting in a single emission line of $E_\gamma\simeq 14.4$~keV, an isomer shift with respect to $\alpha$-Fe of $-5~\mathrm{neV}$ and a linewidth of $7$~neV ($1.7~\mathrm{MHz}$), slightly broader than the natural linewidth of the nuclear state. The absorption spectrum of the ${}^{57}\mathrm{Fe}$ film is scanned by varying the Doppler shift between the source and absorber. The highest contrast of the absorption resonances is about $3\%$, so that the count rate is dominated by non-resonant $\gamma$s ($14.4~\mathrm{keV}$ and Compton scattering of $122~\mathrm{keV}$). A tungsten collimator is used to increase the contrast of the resonances, resulting in a count rate of approximately 2~kHz, consistent with the internal conversion coefficient and solid angle defined by the collimator~\cite{appendix}. For each spectrum collected, about 10 days of data are required to sufficiently suppress Poisson fluctuations.
The amplitude of the SAW generated by the coupler is given by $A = C_\perp\, \eta\sqrt{P_\mathrm{in}}$, where $P_\mathrm{in}$ is the applied RF power, $\eta = 0.35$ is the overall electro-acoustic attenuation~\cite{appendix}, and $C_\perp$ is the out-of-plane displacement conversion constant (in units of $\mathrm{m/\sqrt{W}}$) that depends on the mechanical properties of the substrate and the width of the IDTs. The attenuation factor $\eta$ accounts for losses due to electrical reflection, device geometry, acoustic attenuation, and other non-idealities. The SAW displacement modulates the nuclear transition energy via the Doppler effect. When the SAW is driven at an angular frequency $\Omega_\mathrm{saw}$, the nuclear levels are periodically shifted, resulting in sidebands spaced by $\Omega_\mathrm{saw}$ (Figure~\ref{fig:figure1}b-c). For a pure traveling surface acoustic wave, the resulting absorption spectrum is given by\cite{aivazyan_determination_1974,yamashita_measurement_2024,appendix}:
\begin{equation}
S(\Omega) = \sum_{i} \sum_{n=-\infty}^{\infty} \frac{s_i\, J_n^2(k_0A)}{\left(\Omega - \Omega_i - n  \Omega_\mathrm{saw}\right)^2 + \left(\frac{\Gamma}{2}\right)^2},
\label{eq:spectrum}
\end{equation}
where $k_0$ is the photon wavenumber, and $\Omega_i$ and $s_i$ are the angular frequency and relative intensity of the $i$-th hyperfine transition, $J_n(k_0A)$ denotes the $n$-th order Bessel function of the first kind, and $\Gamma$ is the transmission linewidth in units of angular frequency, given by the sum of the linewidths of the source and absorber.
\begin{figure}[h!]
    \centering
    \includegraphics[width=8.6cm]{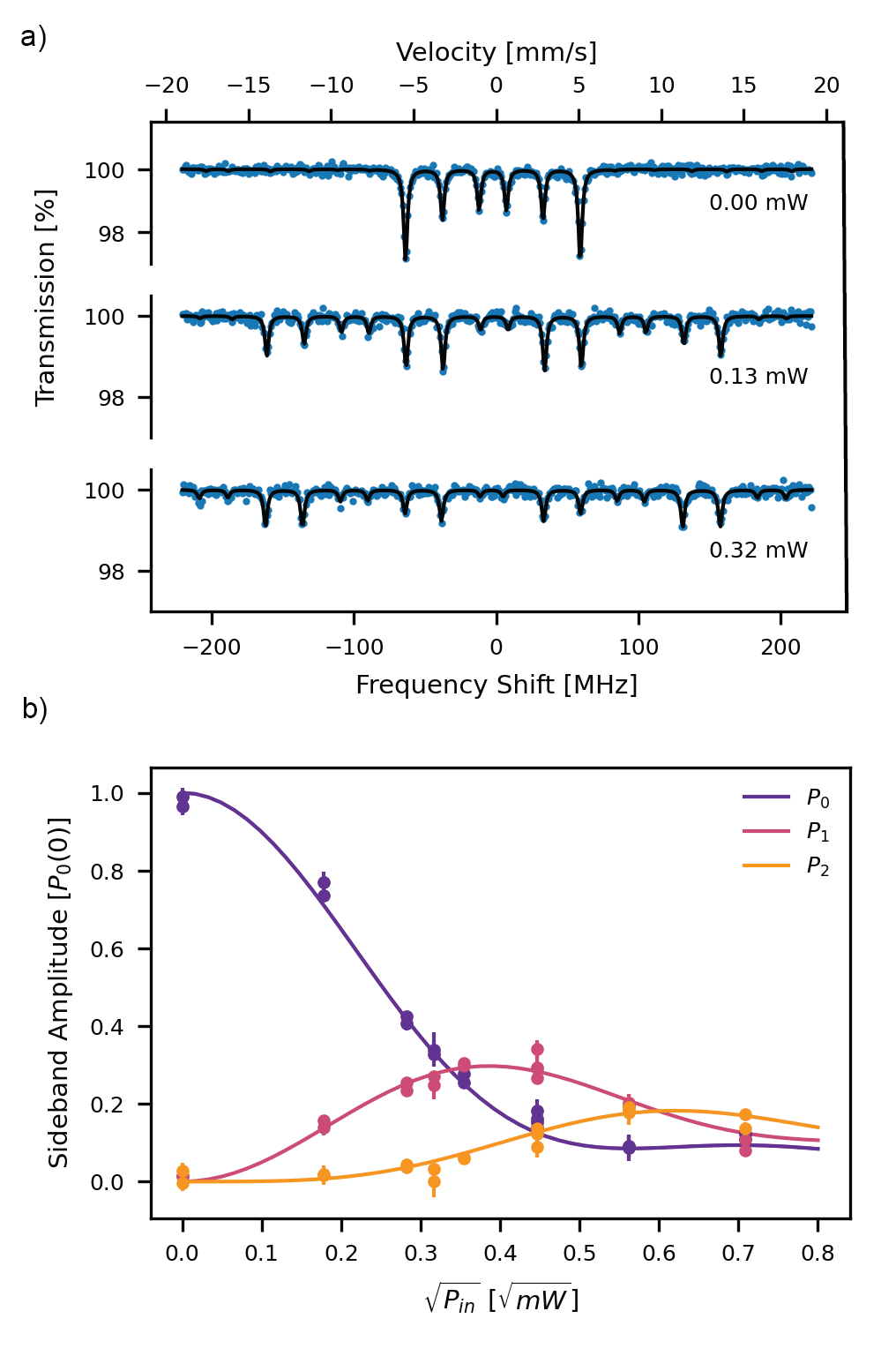}
    \caption{a) Mössbauer absorption spectra of the device with no SAW excitation and two cases of increasing SAW drive power. As the modulation strength grows, higher-order sidebands emerge while amplitudes decrease to maintain a constant overall integral over the absorption spectrum. Continuous lines, overlaying the data points, represent fits to the data~\cite{appendix}. b) Reconstructed amplitudes of the first three sidebands accessible within the available velocity range, as a function of the applied SAW drive power. The solid lines are a global fit to all three datasets using a model based on the squared Bessel functions integrated over the SAW amplitude distribution. The two fit parameters are the modulation scaling factor $C_\perp$ and the reflection coefficient $\alpha$, which encodes the degree of standing-wave formation due to SAW reflections.}
    \label{fig:figure3}
\end{figure}
The relative prominence of the side bands depends on the applied SAW power, which is varied in the experiment to characterize the device. For each setting a spectrum is acquired by scanning the source velocity from $-19 < v <+19 \,\mathrm{mm/s}$ under constant acceleration. Within this velocity range each spectrum reveals up to 18 visible absorption peaks, as shown in Figure~\ref{fig:figure3}a, which reduce, due to symmetry and the known hyperfine structure, to 7 distinct features with characteristic amplitudes which follow combinations of squared Bessel functions $J_n^2(m)$ up to order $n = 2$. Notably, the natural Zeeman splitting in $\alpha$-Fe ($24.4$~MHz) happens to be one fourth of the SAW modulation frequency ($97.9$~MHz), resulting in spectral overlap between certain sidebands and the unshifted hyperfine lines, within our instrumental resolution. While this overlap has no consequence for our experiment, which employs an incoherent source, synchrotron-based excitation would lead to interference between the unmodulated transitions and the overlapping sidebands.

To extract quantitative information, each spectrum is fit using a sum of Lorentzian peaks overlaid on a fifth-order polynomial baseline, accounting for the change in solid angle in the scan, non-uniform channel sampling, and slow drifts in detector response. The fit yields precise absorption peak locations that allow the calibration of the Doppler velocity scale over the full $\pm19~\mathrm{mm/s}$ range~\cite{appendix}.
Each spectrum is then normalized by its polynomial baseline, resulting in the spectra in Figure~\ref{fig:figure3}a, and the area under each peak is integrated. From the resulting integrated spectra, the area of each sideband can be directly evaluated. Combined with knowledge of the power delivered to the SAW coupler, this enables the reconstruction of the modulation indices $k_0A(P_{in})$.  The residual reflection from the output transducer results in a spatially inhomogeneous SAW amplitude along the propagation direction which can then be expressed as $A(x) = A_0 \sqrt{1 + \alpha^2 + 2\alpha \cos( k_\mathrm{saw} x )}$, where $A_0$ is the amplitude of the forward-propagating wave, $\alpha$ is the coherent reflection coefficient, and $k_\mathrm{saw}$ is the SAW wavenumber. To account for this effect, the model described in Eq.\ref{eq:spectrum} is integrated over the amplitude distribution, yielding a modified prediction for the sideband structure~\cite{appendix, aivazyan_determination_1974, yamashita_measurement_2024}. Both the out-of-plane displacement conversion constant
$C_\perp = (5.35\pm 0.37) \cdot 10^{-9}~\mathrm{m/\sqrt{W}}$ and the reflected wave fraction $\alpha = 0.36\pm 0.02 $ are extracted by fitting this extended model (Figure~\ref{fig:figure3}b). These values are in excellent agreement with electromechanical calculations for an ST-cut quartz substrate supporting a Rayleigh-type surface acoustic wave, based on the known elastic constants of quartz~\cite{appendix, slobodnik_surface_1976, coquin_analysis_1967}, and are consistent with near-unity transfer of the SAW displacement to the $^{57}$Fe film within experimental uncertainty.

\section*{Discussion and Outlook}
The results presented here demonstrate the successful dynamic modulation of \M nuclear energy levels using  $97.9$ MHz surface acoustic waves. This high frequency control, both in speed and in phase coherence, establishes a new regime of acoustic–nuclear coupling. Unlike lower-frequency or quasi-static piezo-driven systems, SAWs offer the advantage of continuous, propagating phase control at GHz bandwidths. In addition, this is achieved in a fully monolithic device, enabling scalable fabrication with the robustness and flexibility of solid-state architectures.
Looking forward, we envision several developments of this technology. First, SAW transducers can be engineered to allow nearly arbitrary mechanical drive, enabled by multi-harmonic excitation~\cite{stejskal_toward_2025} or by chirped designs~\cite{fall_generation_2017}. Combined with synchrotron radiation (with X-ray spot-sizes smaller than the SAW wavelength) such devices would allow fast, coherent programmable modulation of nuclear transitions, extending the reach of time-domain control protocols in nuclear quantum optics, including fast phase-jump and interferometric schemes~\cite{vagizov_coherent_2014,heeg_spectral_2017,gerharz_dark-fringe_2025}.
Beyond waveform control, the broad tunability of the SAW frequency and amplitude provides a stable and precisely controllable Doppler shift of thin Mössbauer absorbers. This capability could significantly enhance the dynamic range of Mössbauer spectrometers, enabling high-precision spectroscopy at much higher Doppler shifts.
A complementary direction is to couple the nuclear resonance to a surface acoustic wave resonator operated at cryogenic temperatures~\cite{schuetz_universal_2015,manenti_surface_2016,satzinger_quantum_2018,patel_surface_2024}, opening a path toward optomechanical experiments in the X-ray regime~\cite{liao_optomechanically_2017}. While single-phonon, single-photon strong coupling is not currently accessible with state-of-the-art SAW resonators, such devices in the quantum regime can nevertheless serve as high-fidelity, quantum-calibrated probes of the local mechanical and thermal environment governing the nuclear states of \M films.
Finally, the ability to couple surface acoustic waves to nuclear transitions may offer a promising route for the dynamic control of nuclear clocks~\cite{shvydko_resonant_2023, tiedau_laser_2024, elwell_laser_2024}, in particular the $^{229}\mathrm{Th}$ isomeric transition which was recently observed in a $^{229}\mathrm{ThF}_4$ thin film~\cite{zhang_229thf4_2024}, potentially enabling mechanical modulation of ultra-stable, nuclear optical transitions, and establishing a new interface between solid-state mechanics and nuclear quantum metrology.

\section*{Acknowledgments}
We thank Prof.~D.~Ryan (McGill), for early advice on \M spectroscopy and Prof.~D.~Goldhaber-Gordon (Stanford) for advice on the design and fabrication of the SAW devices. This work was supported by the Air Force Office of Scientific Research under award number FA9550-22-1-0439, and by the "Table-top experiments for fundamental physics" program, sponsored by the Gordon and Betty Moore Foundation, Simons Foundation, Alfred P. Sloan Foundation, and John Templeton Foundation. Sample fabrication was partially supported by the U.S. Department of Energy, Office of Science,
Basic Energy Sciences, Materials Sciences and Engineering Division, under Contract No. DE-AC02-76SF00515. Part of this work was performed at the Stanford Nano Shared Facilities (SNSF) and the Stanford Nanofabrication
Facility (SNF), both of which are supported by the National Science Foundation under Award No. ECCS-2026822.

\FloatBarrier
\let\oldaddcontentsline\addcontentsline
\renewcommand{\addcontentsline}[3]{}
\bibliographystyle{naturemag}
\bibliography{Bib}
\let\addcontentsline\oldaddcontentsline

\clearpage
\onecolumngrid

\begin{center}
\large\textbf{\uppercase{Appendix}}
\end{center}


\renewcommand{\thetable}{A\arabic{table}}  
\renewcommand{\thefigure}{A\arabic{figure}} 
\renewcommand{\theHfigure}{A\arabic{figure}}
\renewcommand{\theequation}{A\arabic{equation}} 
\setcounter{figure}{0}
\setcounter{equation}{0}
\setcounter{section}{0}
\setcounter{table}{0}

\maketitle
\vspace{2,5cm}
\tableofcontents
\vspace{2,5cm}

\section{Theory}
\label{sec:theory}

\subsection{Quantum model for \M absorption under surface acoustic wave modulation}

\begin{figure}[h]
    \centering
    \includegraphics[width=5cm]{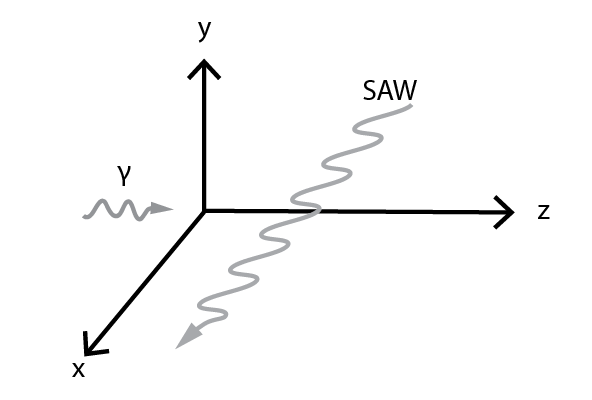}
    \caption{Definition of the coordinate system: the $\gamma$-rays travel along $z$, encountering an ${}^{57}$~Fe film parellel to the $x-y$ plane and driven by a SAW travelling along $x$.}
    \label{fig:coordinates}
\end{figure}

The interaction between the nuclear transition and the phonon field can be modeled via the coupling of the quantized $\gamma$-ray field to a phonon-dressed nuclear dipole operator. The phonon dressing accounts for lattice vibrations at the nuclear site, while the SAW drive introduces a time-periodic modulation of the transition, resulting in Floquet sidebands. We define coordinates with $\gamma$-rays traveling along the z direction, the SAW propagating along x, and the SAW-induced nuclear displacement along z (see Figure~\ref{fig:coordinates}.
To derive the interaction of the SAW with the nucleus, we start from the dipole interaction with the electric field \cite{Scully_Zubairy_1997}
\begin{equation}
    \hat H_{int} = -\hat{d}\cdot \hat{E},
    \label{dipoleint}
\end{equation}
where $\hat{d} =d(\hat\sigma_+e^{i\omega_a t} +\hat\sigma_-~e^{-i\omega_a t})$ is the dipole operator with $\hat\sigma_+ = \ket{e}\bra{g} ~$ and $\hat \sigma_- = \ket{g} \bra{e}$, describes the lowering and raising operator and $\omega_a$ the transition frequency of the absorber. The quantized electric field associated with the $\gamma$-rays emitted by the radioactive source is given by
\begin{equation}
\hat{E} = \epsilon\left(\hat a e^{-i\omega_{0} t + i k_0 \hat{z}} + \hat a^\dagger e^{i\omega_{0} t - i k_0 \hat{z}}\right),
\label{Efield}
\end{equation}
where  $a$ and $a^\dagger$ are the photon annihilation and creation operators, respectively, $\omega_0$ is the $\gamma$-ray frequency, and $k_0$ is the associated wave vector. $\epsilon = \sqrt{\frac{\hbar \omega_0}{2 \epsilon_0 V}}$ denotes the coupling strength of the electromagnetic field, with $V$ the mode volume and $\epsilon_0$ is the vacuum permittivity. The position operator $\hat{z}$ reflects the spatial dependence of the field and will later incorporate modifications due to mechanical motion induced by the surface acoustic waves \cite{Scully_Zubairy_1997}. Plugging the dipole operator and the quantized electric field \ref{Efield} into the interaction Hamiltonian \ref{dipoleint}, results in
\begin{equation}
    \hat H_{int}= -g(\hat \sigma_+e^{i\omega_a t} +\hat \sigma_-~e^{-i\omega_a t})\left[\hat ae^{-i\omega_{0} t}e^{ik_0\hat{z}}+\hat a^\dagger e^{i\omega_{0} t}e^{-ik_0\hat{z}}\right], 
\end{equation}
where we have defined the coupling constant $g = \epsilon\cdot d$. The expansion produces fast terms $e^{\pm i(\omega_a+  \omega_0) t}$, which in the rotating-wave approximation average to zero. 
Defining the detuning between the incoming photons and the nuclear resonance as  $\Omega = \omega_a - \omega_0$, the interaction Hamiltonian becomes:
\begin{equation}
\hat H_{int} = -g\left[\hat \sigma_+ a e^{i\Omega t} e^{ik_0 \hat{z}} +\hat \sigma_- a^\dagger e^{-i\Omega t} e^{-ik_0 \hat{z}}\right]  -i\frac{\Gamma}{2}\ket{e}\bra{e}.
\end{equation}
where the non-Hermitian term \(-i\frac{\Gamma}{2}\ket{e}\!\bra{e}\) is introduced to incorporate spontaneous decay via the Weisskopf–Wigner approximation \cite{weisskopf_berechnung_1930,kuchment_floquet_1982,salkola_time-dependent_1990}. The displacement of the nucleus due to the phonon field, denoted by $\hat{z}$, can be described as the sum of an average coherent term and a fluctuating term:~\cite{aspelmeyer_cavity_2014}
\begin{equation}
\hat{z}(t) = \bar z(t) + \delta \hat z = A \cos(\Omega_{\text{saw}} t) + z_{\text{zpf}}(\hat b + \hat b^\dagger).
\end{equation}
where, A is the amplitude of the nucleus driven by the surface acoustic wave, $\Omega_\mathrm{saw}$ frequency of the driven SAW, $z_\mathrm{zpf} = \sqrt{\frac{\hbar}{2\rho v_\mathrm{s}A_\mathrm{saw}}}$ the zero-point fluctuation of that SAW mode ($v_s$: speed of sound in the substrate, $\rho$: mass density of quartz, $A_\mathrm{saw}$: mode area of the SAW mode) \cite{schuetz_universal_2015}. Note that in principle the fluctuating part of the displacement operator is spread over all frequencies supported by the material. For simplicity the analysis is limited to the single mechanical mode at $\Omega_\mathrm{saw}$ defined by the creation and annihilation operators $\hat b^\dagger,\,\hat b$. Substituting the nuclear displacement $\hat z(t)$ and the decay term into the interaction Hamiltonian yields the full time-dependent Hamiltonian that captures both the classical drive and quantum fluctuations of the SAW:
\begin{equation}
    \hat H_{int}= -g\left[\hat a e^{-i\Omega t} e^{ik_0 z_\mathrm{zpf}(\hat b+\hat b^\dagger)}e^{iA\cos(\Omega_\mathrm{saw}t)}\hat\sigma_+ +h.c.\right] -i\frac{\Gamma}{2}\ket{e}\bra{e}.
\end{equation}
The interaction Hamiltonian is periodic in time and can be treated as a Floquet Hamiltonian. According to Floquet's theorem, such a Hamiltonian can be expanded in terms of its Fourier components. Given the structure of the interaction, we can apply the Jacobi–Anger expansion \cite{bessel_theory}, $e^{izcos(\theta)}=\sum_{n=-\infty}^{\infty}i^n J_n(z) e^{in\theta}$,
to rewrite the Hamiltonian as a sum over harmonic components. These are known as Floquet sidebands, weighted by Bessel functions of the first kind, $J_n(z)$. This decomposition allows to express the interaction as a sum over discrete frequency components indexed by $n$, each corresponding to a sideband in the modulated spectrum. Then the interaction Hamiltonian takes the form of
\begin{equation}
    \hat H_{int}= -g\sum_n i^n J_n(k_{0}A) \left[\hat a\hat \sigma_+e^{-i(\Omega-n\Omega_\mathrm{saw})t}e^{ik_{0}z_\mathrm{zpf}(\hat b+\hat b^\dagger)}  + \hat a^\dagger \hat \sigma_- e^{i(\Omega-n\Omega_\mathrm{saw})t}e^{-ik_{0}z_\mathrm{zpf}(\hat b+\hat b^\dagger)} \right]-i\frac{\Gamma}{2}\ket{e}\bra{e},
\end{equation}
showing the emergence of $n$ periodic sidebands. The periodic time dependence introduced by the SAW results in a Floquet Hamiltonian, with quasi-energy sidebands from the modulation of the nuclear state.
To obtain the experimentally observable quantity, the absorption spectrum, it is necessary to calculate the transition amplitudes between the relevant quantum states. For each nucleus, the emission process corresponds to a transition from the excited state $\ket{e}$ to the ground state $\ket{g}$, accompanied by the emission of a photon. In addition, the phonon occupation number changes during the transition ($m\rightarrow m^\prime$). The corresponding transition amplitude is then given by
\begin{equation}
 M_{i \rightarrow f}(t)= -g \sum_n (-i)^n J_n(k_{0} A)\, c\, e^{i(\Omega - n\Omega_\mathrm{saw})t} e^{-\Gamma t / 2}
\end{equation}
The phonon matrix elements are given as $c = \bra{m'} e^{-ik_0 z_\mathrm{zpf}(\hat b +\hat b^\dagger)} \ket{m} = 1 + O(\sqrt{m}k_0 z_\mathrm{zpf})$, capturing the influence of lattice vibrations through transitions in the phonon Fock space. However, at room temperature ($\bar m= k_\mathrm{B}T/\hbar\Omega\mathrm{saw}$) the system operates in the Lamb–Dicke regime, where  $\bar m (k_0 z_\mathrm{zpf})^2<<1$, and interactions with thermal phonon fluctuations become negligible~\cite{cohen-tannoudji_atom-photon_1998}. Accounting for Zeeman splitting into multiple hyperfine transitions with amplitudes $s_i$ and frequency shifts $\Omega_i$ the absorption spectrum is:
\begin{equation}
S(\Omega)  =\frac{|g|^2}{2\pi}\sum_n\sum_i \frac{J_n^2(k_{0}A) s_i}{\left(\Omega - n\Omega_\mathrm{saw}-\Omega_i \right)^2 + \left(\frac{\Gamma}{2}\right)^2}.
\end{equation}
This expression captures the signature of Floquet-dressed nuclear states, manifesting as energy sidebands spaced by integer multiples of $\Omega_\mathrm{saw}$ in the emission spectrum.

\subsection{Correction for a non-uniform drive amplitude}

Because the readout coupler partially reflects the SAW, the displacement in the ${}^{57}$Fe  film is a coherent superposition of a forward-traveling wave and a weaker reflected component. Denoting the reflected-to-incident amplitude ratio by  $\alpha$, and assuming a coherent phonon state, the coherent displacement can be written as:
\begin{equation}
    \bar z(t, x) = A_0 e^{i\Omega_\mathrm{saw} t}[(1-\alpha)e^{-ik_\mathrm{saw}x}+2\alpha \cos(k_\mathrm{saw}x)].
\end{equation}
which can be rewritten as:
\begin{equation}
    \bar z(t,x) =  A_0 f(x) \cos(\Omega_\mathrm{saw}t +\phi^\prime) 
    \label{displacement}
\end{equation}
with $f(x) =\sqrt{1+\alpha^2+2\alpha \cos(2k_\mathrm{saw}x)}$ . In the present experimental implementation, there is no spatial resolution. Consequently, the measured spectrum corresponds to a spatial average over the standing-wave amplitude~\cite{aivazyan_determination_1974,yamashita_measurement_2024}. To obtain the corresponding probability distribution for the local amplitude, the wave must be integrated over the full interaction length (L=4~mm). The resulting normalized amplitude distribution is given by:
\begin{equation}
    p(y) = \frac{1}{L} \int_0^L dx\, \delta\left(y - f(x)\right)
    = \frac{2y}{\pi\sqrt{4\alpha^2 - (y^2 - 1 - \alpha^2)^2}}.
\end{equation}
To determine the actual power $P_n$ in each sideband $n$, the corresponding Bessel function must be integrated over all possible amplitudes, weighted by their probability distribution:
\begin{equation}
    P_n(k_0A_0)= \int_{1-\alpha}^{1+\alpha}dy~p(y) J_n^2(k_0 A_0 y).
\end{equation}
This integral is evaluated numerically, and the resulting values are used as input to fit the experimental data. The spectrum is finally:
\begin{equation}
    S(\Omega) = \sum_{i} \sum_{n} \frac{s_i\,  P_n(k_0A_0)}{\left(\Omega-\Omega_{i} - n\Omega_\mathrm{saw}  \right)^2 + \left(\frac{\Gamma}{2}\right)^2}.
\label{eq:spectrum}
\end{equation}

\subsection{Quartz surface acoustic wave motion calculation}
\begin{figure}[h!]
\includegraphics[width=18cm]{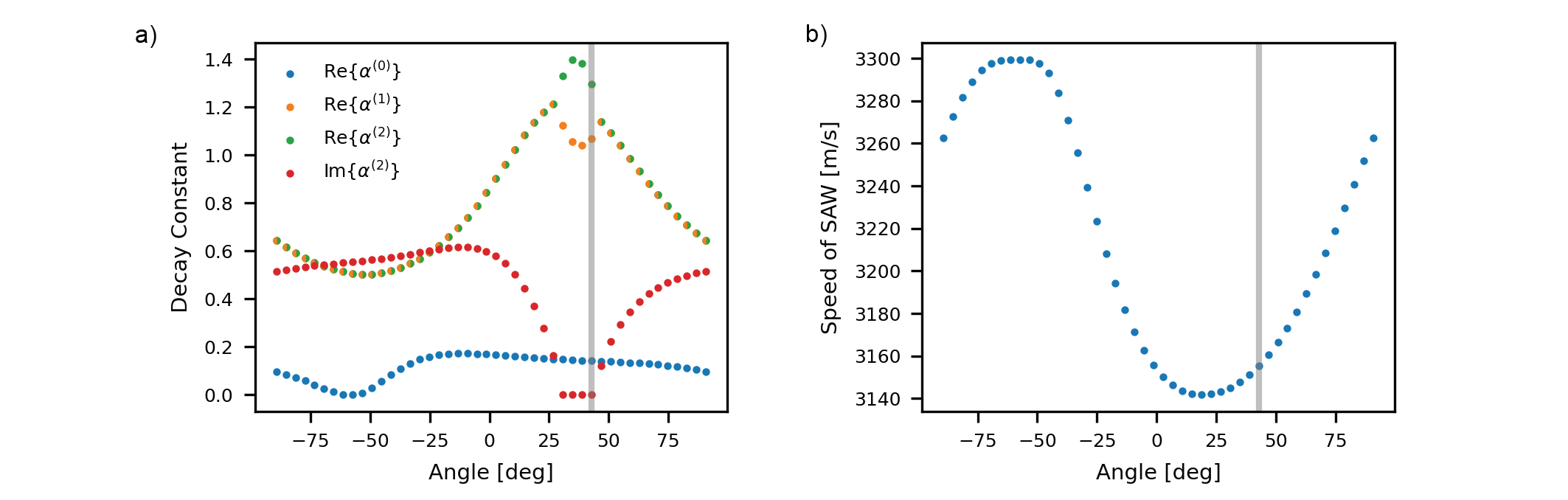}
\caption{Numerical calculation of the decay constants (a) and the speed of the SAW (b) versus the cut angle of the Y-cut quartz. These figures reproduce figures 3 and 4 in ref.~\cite{coquin_analysis_1967}. The gray lines indicate the angle of ST-cut quartz ($+42.75^{\circ}$).}
\label{fig:elastic}
\end{figure}

\FloatBarrier
We calculate the expected out-of-plane displacement conversion constant $C_\perp^{theory}$ by finding surface acoustic wave solutions that satisfy the elastic equation of motion in quartz and respect the boundary conditions of the problem. The equation of mechanical motion in quartz is written in terms of the mechanical displacement $u_i$ and the elastic tensor $c_{ijkl}$, where the indices label the direction (1, 2, 3 $\rightarrow$ x, y, z) and indices after a comma indicate a spatial derivative:
\begin{equation}
    \rho \ddot{u}_{i} = c_{ijkl} u_{k,lj}.
    \label{equation_of_motion}
\end{equation}
As the quartz is weakly piezoelectric, the piezoelectric contribution to the equation of motion can be ignored~\cite{coquin_analysis_1967}. The boundary conditions of the problem are such that the surface acoustic wave must be confined to the surface of the quartz:
 \begin{equation}
    \lim_{z\rightarrow \infty}u_i = 0,
    \label{surface_wave_bc}
\end{equation}
and that the surface of the quartz remains stress free:
\begin{equation}
    T_{3j} = c_{3jkl}u_{k,l}=0 ~ \mathrm{when} ~ z=0.
    \label{stress_free_bc}
\end{equation}
The mechanical motion in quartz is anisotropic, so the orientation of the crystal cut and the propagation direction of the surface acoustic wave are important. We are interested in SAW modes which propagate along the crystal X-axis of quartz; this axis corresponds to a digonal symmetry in the quartz. The mechanical parameters were calculated for different rotated Y-cuts of quartz by appropriately rotating the quartz elastic tensor. The following ansatz is used to find solutions to the mechanical problem:
\begin{equation}
    u_i = C_{i}^{(j)} e^{-\alpha^{(j)} \Omega_{\mathrm{saw}} z/v_s}e^{i\Omega_{\mathrm{saw}} (t-x/v_s)}
    \label{ansatz}
\end{equation}
where $\alpha^{(j)}$ ($j = 1,2,3$) are the decay constants, $v_s$ is the speed of the SAW, and $C_{i}^{(j)}$ is the amplitude of displacement in the the $i$-th direction due to the $j$-th decay constant. We numerically solve the mechanical problem for different angles, reproducing the results from Coquin and Tiersten~\cite{coquin_analysis_1967}(Figure \ref{fig:elastic}).

\FloatBarrier
\FloatBarrier
    
From the calculated values of $C_i^{(j)}$,  $\alpha^{(j)}$ and $v_s$ we compute the power in the surface acoustic wave defined by:
\begin{equation}
    P_x = -\frac{1}{4}Re\int_0^\infty c_{1jkl} \left (u_{k,l} + u_{l,k} \right ) {\dot{u}_j}^*~dz
    \label{power_integral}
\end{equation}
and the amplitude of the surface acoustic wave:
\begin{equation}
    A = \left | \sum_{j=1}^3 C_{3}^{(j)} \right |
    \label{amplitude_sum}.
\end{equation}
Finally we compute the expected out-of-plane displacement conversion constant for the $+42.75^{\circ}$ rotated Y-cut of quartz:
\begin{equation}
    C_\perp^{theory} = \frac{A}{\sqrt{w~P_x}} = 5.511 \cdot 10^{-9}~\mathrm{m}/\sqrt{\mathrm{W}}
    \label{theory_C}
\end{equation}
where $w = 970 ~\upmu\mathrm{m}$ is the width of the SAW device.

\section{Fabrication}

\subsection{Enriched $^{57}\mathrm{Fe}$ deposition}
In order to deposit Enriched $^{57}\mathrm{Fe}$ film in a clean environment we designed and built a dedicated thermal evaporator. The growth chamber base pressure is $9.5 \cdot 10^{-11}$~Torr. The chamber includes a load-lock, a rotating substrate holder, a substrate heater, a ${}^{57}$Fe effusion cell, and a quartz crystal monitor. UHV pressure measurements were performed with a Bayard-Alpert gauge and the results agree with the pressure inferred from the current consumption of the Agilent $500$~L/s noble diode ion pump.
A $2$" O.D., $500$~$\upmu$m thick SSP ST-Cut quartz wafer (MSE supplies). is placed in the substrate holder of the deposition chamber, set to rotate at approximately 12.5 rpm, and the substrate heater is set at $400~^{\circ}$C for for $6$ days.
After wafer baking, the crucible in the ${}^{57}$Fe effusion cell is heated to $1270~^{\circ}$C for approximately $5$ minutes. The crucible temperature is then lowered to $1220~^{\circ}$C for a deposition time of $32$ hours and $43$ minutes. During deposition, the pressure in the growth chamber is around $9.5\cdot 10^{-10}$~Torr, and the average temperature of the substrate holder is approximately $75~^{\circ}$C.

\subsection{SAW transducers design and fabrication}

A delay-line SAW device with finger spacing of $\lambda/4$ is designed in KLayout. Each coupler (interdigital transducer, IDT) comprises 800 fingers (200 sets of 4 finger periods) at a design wavelength of $32~\upmu\mathrm{m}$. The IDT width is $970~\upmu\mathrm{m}$, the distance between transducer centers is $12~\mathrm{mm}$, and the fingers are connected by a $250~\upmu\mathrm{m}$-wide bus bar.
Using a custom vacuum deposition apparatus, a $300~\mathrm{nm}$ film of enriched ${}^{57}\mathrm{Fe}$ is deposited on an ST-cut quartz wafer. A $1.6~\upmu\mathrm{m}$ layer of AZ-1512 photoresist is then spun and baked at $95~^{\circ}\mathrm{C}$ for $5~\mathrm{min}$. The wafer is exposed in a MicroWriter ML3 (nominal $\lambda=385~\mathrm{nm}$) at $0.6~\upmu\mathrm{m}$ resolution to define an array of 11 devices, with doses spanning $110$–$170~\mathrm{mJ\,cm^{-2}}$. After development in Microposit MF-319, the pattern is is wet etched for $40~\mathrm{s}$ in $4~\mathrm{M}$ HCl. The wafer is rinsed in deionized (DI) water, dried with nitrogen, and descumed in a weak oxygen plasma. A $60$~\r{A} Ti adhesion layer and $600$\r{A} Au are e-beam evaporated, and liftoff is achieved by gentle sonication in acetone. Post-liftoff, the wafer is cleaned in isopropanol (IPA) and nitrogen-dried.
After fabrication of the SAW transducers a second round of lithography is executed to define the ${}^{57}\mathrm{Fe}$ interaction window.
A $900\times 4000~\upmu\mathrm{m}^2$ rectangle centered on each SAW device is defined on a new layer of photoresit with a resolution of $1.0~\upmu\mathrm{m}$ and $150~\mathrm{mJ\,cm^{-2}}$ dose. After development and cleaning $60$~\r{A} Ti and $60$~\r{A} Au are deposited in the  e-beam evaporator. Liftoff is achieved by gentle sonication in acetone. Finally the wafer is protected with an overcoat of AZ-1512.
The wafer is diced into individual chips, which are optically inspected for shorts and defects. A representative device is selected, solvent-cleaned (acetone, then IPA) and nitrogen-dried, then briefly etched in dilute $\mathrm{FeCl}_3$ until the majority of the quartz substrate becomes optically transparent, leaving a film of ${}^{57}\mathrm{Fe}$ at the center of the device. After a final DI rinse, IPA clean, and nitrogen dry, the chip is mounted to a carrier PCB with double-sided tape, and the IDT bus bars are wire-bonded to the PCB. Small pieces of Kapton tape are attached to the quartz ends to suppress edge reflections.

\section{Experiment characterization}

\FloatBarrier
\subsection{SAW electromechanical characterization}
\label{sec:s-parameters}

The transducer is modeled as an acoustic conductance ($G_\mathrm a$) and an acoustic susceptance ($B_\mathrm a$) in parallel with a capacitor $C_\mathrm T$ and in series with a resistor $R_ \mathrm L$. The form of the acoustic admittance is taken from the cross-field model described by Slobodnik~\cite{slobodnik_surface_1976}, and the capacitor and resistor describe the non-acoustic electrical response of the SAW transducer.
Equation \ref{eq:s11s12} describe the s-parameters for a single transducer (i.e. the ratio between the source and the SAW amplitude).
\begin{figure}
    \centering
    \includegraphics[width=14cm]{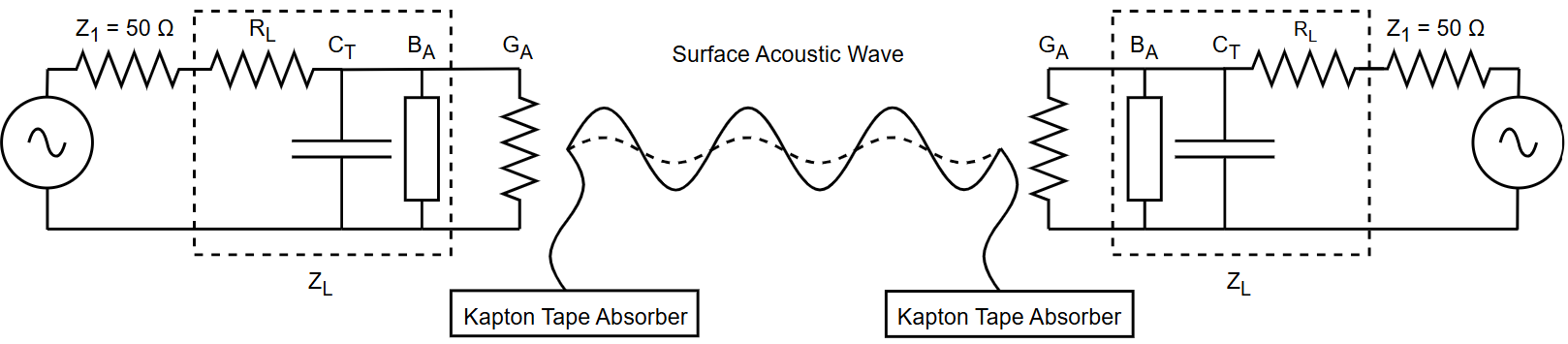}
    \caption{Model of the surface acoustic wave transducer device.}
    \label{fig:figureSI_2}
\end{figure}
Defining the normalized detuning as  $x = N\pi(\Omega-\Omega_\mathrm{saw})/\Omega_\mathrm{saw}$ ($N$: number of finger periods), the acoustic admittance and electrical capacitance of the transducer can be written as:
\begin{equation}
\begin{aligned}
G_{\mathrm a}(x) &= \frac{4}{\pi}\,k^2\,\Omega_\mathrm{saw}\,C_{\mathrm{FF}}\,N^{\,2}
\left(\frac{\sin x}{x}\right)^{\!2},\\[4pt]
B_{\mathrm a}(x) &= \frac{4}{\pi}\,k^2\,\Omega_\mathrm{saw}\,C_{\mathrm{FF}}\,N^{\,2}
\left(\frac{\sin(2x)-2x}{2x^{2}}\right),\\ 
C_\mathrm T &= N\,W\,C_{\mathrm{FF}},
\end{aligned}
\end{equation}
 where \(k^2\) is the electromechanical coupling factor, \(C_{\mathrm{FF}}\) is the capacitance per finger period per transducer width, \(W\) is the width of the IDT aperture and \(C_{\mathrm T}\) is the total IDT capacitance. For a single transducer, the source impedance at port 1 is \(Z_1=50~\Omega\), the impedance at port 2 is \(Z_2=1/G_{\mathrm a}(x)\) and the impedance of the shunt is calculated as \(Z_{\mathrm L}=R_{\mathrm L}+\big(\mathrm j\,\omega C_{\mathrm T}+\mathrm j\,B_{\mathrm a}(x)\big)^{-1}\).
Following the scattering parameter formalism~\cite{kurokawa_power_1965} the s-parameters for each transducer is:
\begin{equation}
\begin{aligned}
S_{11}^\prime &= \frac{Z_{\mathrm L} Z_2 - Z_1^{\ast}\!\left(Z_{\mathrm L}+Z_2\right)} {Z_{\mathrm L} Z_2 + Z_1\!\left(Z_{\mathrm L}+Z_2\right)}, \qquad 
S_{22}^\prime = \frac{Z_{\mathrm L} Z_1 - Z_2^{\ast}\!\left(Z_{\mathrm L}+Z_1\right)} {Z_{\mathrm L} Z_1 + Z_2\!\left(Z_{\mathrm L}+Z_1\right)},
\\[6pt]
S_{12}^\prime &= S_{21}^\prime = \frac{\sqrt{Re(Z_1)}}{\sqrt{Re(Z_2)}}\, 
\frac{Z_{\mathrm L}\!\left(Z_2 + Z_2^{\ast}\right)}
{Z_1 Z_2 + Z_{\mathrm L}\!\left(Z_1 + Z_2\right)}.
\label{eq:s11s12}
\end{aligned}
\end{equation}
The s-parameters for the entire device, which can be directly measured (Figure \ref{fig:s11}) can be approximated to:
\begin{equation}
S_{11} \approx S_{11}^\prime, \qquad S_{12} \approx \left(S_{12}^\prime \right )^2\frac{\eta_\mathrm p}{2},
\label{eq:s12}
\end{equation}
where $\eta_ \mathrm p$ models the attenuation of the surface acoustic wave due to propagation and diffraction effects in quartz. For the devices described here, the calculated loss amounts to $-1.68~\mathrm{dB}$\cite{slobodnik_surface_1976}.
The attenuation of the input wave up to the center of the device ($\eta$) and the fraction of the surface acoustic wave that reflects off the second transducer and reaches the center of the SAW device ($\alpha$) are calculated as:
\begin{equation}
    \eta = \sqrt{\frac{\eta_p}{2}} S_{12}^\prime(\Omega_\mathrm{saw}), \qquad \alpha = \frac{\eta_p}{2} S_{22}^\prime(\Omega_\mathrm{saw}).
\label{s_model}
\end{equation}
Fitting the s-parameter equations \ref{s_model} to the measured data in \ref{fig:s11}, we extract an electro-acoustic attenuation coefficient of $\eta = 0.35$ and a surface acoustic wave reflection coefficient of $\alpha = 0.34$.
\begin{figure}[h!]
    \centering
    \includegraphics[width=18cm]{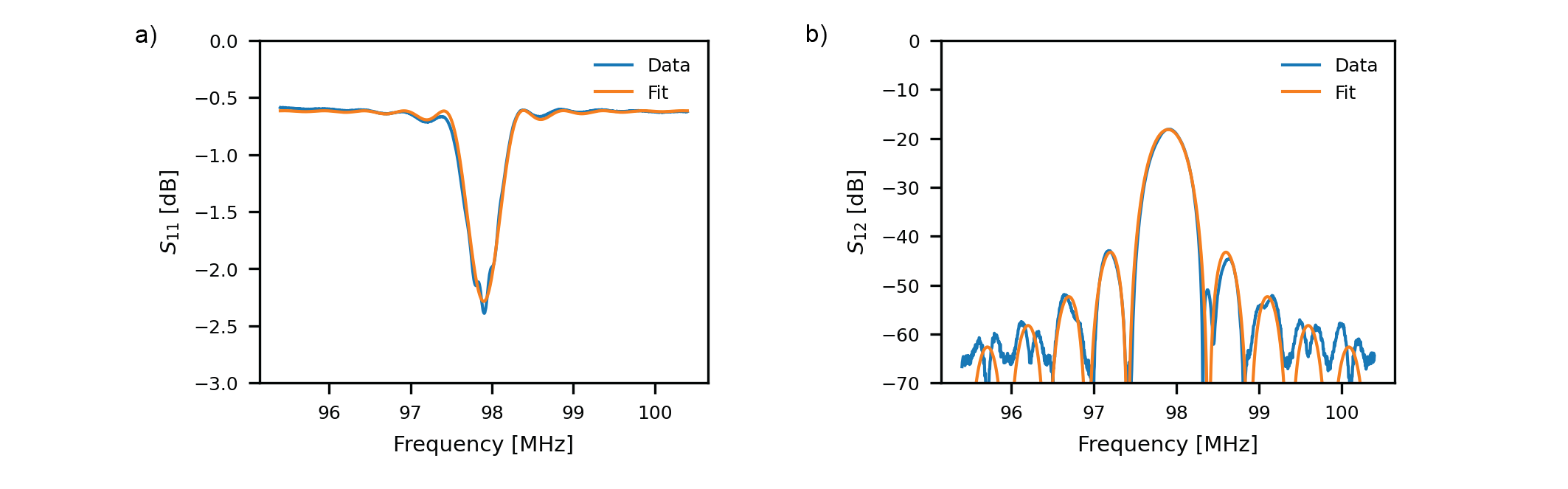}
    \caption{Measurements (blue lines) of the $S_{11}$ (a) and $S_{12}$ (b) , and fit (orange lines) to the model defined in equations~\ref{eq:s11s12}}
    \label{fig:s11}
\end{figure}

\FloatBarrier
\subsection{SEM film characterization}
\begin{figure}[h]
    \centering
    \includegraphics[width=12cm]{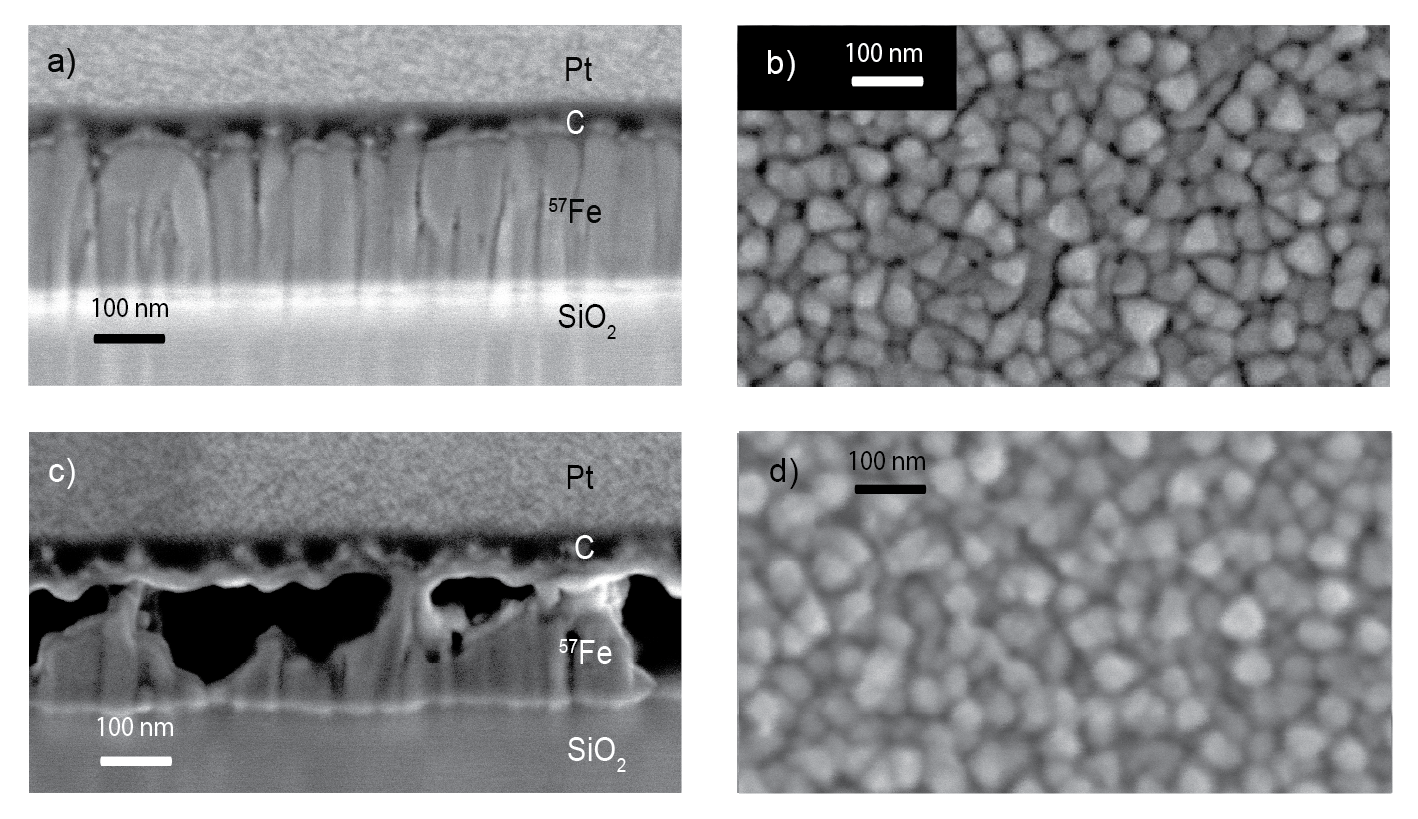}
    \caption{\textbf{SEM of the ${}^{57}$~Fe film.} (a,c) Cross-sections at $52^{\circ}$; (b,d) plan-views. (a,b) Before etch; (c,d) after etch. In (a) the $60$~\r{A} layer of gold is visible between the carbon layer and the iron. Because of the crystalline topography in the iron growth, this gold layer is only partially effective as an etch stop. The over etch produces $\sim300$~nm wide pits (c) that reduce the contrast of the \M lines, but leave the SAW mechanics and Zeeman-line ratios unaffected.}
    \label{fig:SEM}
\end{figure}

The FEI Helios NanoLab 600i DualBeam FIB–SEM was used to characterize the ${}^{57}$Fe films. Two specimens were imaged: (i) a quartz-supported ${}^{57}$Fe film prior to wet etching and (ii) a section taken from a completed SAW device after the final fabrication etch. To enable high-contrast cross-section imaging while minimizing milling-induced damage, we first deposited a carbon cap over a $15~\upmu\mathrm{m}\times 2~\upmu\mathrm{m}$ area by electron-beam–induced deposition (EBID), followed by a $1~\upmu\mathrm{m}$ platinum protective layer by ion-beam deposition. Cross sections were prepared by milling to a depth of $2~\upmu\mathrm{m}$, followed by a cleaning cross section to improve surface quality. Interior views were acquired at a stage tilt of $52^{\circ}$, together with plan-view images of the film surface (Fig.~\ref{fig:SEM}). 
The post-etch cross-section (Fig.~\ref{fig:SEM}c) shows voids created by the etching process due to holes in the etch stop. Because these features are much smaller then the SAW wavelength, $\lambda_\mathrm{saw}=32~\upmu\mathrm{m}$, the device’s mechanical response are unchanged. The effect of the over etch is the reduction by about a factor of 2.2 of the absolute \M contrast due to the reduced ${}^{57}$~Fe filling fraction, which corresponds to an effective thickness of the iron film of about $90$~nm.

\FloatBarrier
\subsection{Efficiency and Loss Budget of the \M Spectrometer}

We characterize the background count rate and detection efficiency of our spectrometer. The total count rate includes not only resonant $14.4$~keV photons but also significant contributions from non-resonant processes. In particular, Compton scattering of the $122$~keV photons from the de-excitation of the $136$~keV state of ${}^{57}$Fe fed by the ${}^{57}$Co EC to the $14.4$~keV \M state, contributes a substantial background component. By using reference targets of various thicknesses and accounting for the known attenuation lengths at different photon energies, we estimate that, off resonance,  for each detected $14.4$~keV photon, there are approximately 0.38 additional counts due to Compton background.

With the SAW-modulated device in place, additional background counts arising from a geometric mismatch between the tungsten collimator and the actual size of the $^{57}$Fe absorber are observed. While the iron film spans $(4\cdot 0.9)~\mathrm{mm}^2$  the collimator aperture measures $4.3\cdot 1.3~\mathrm{mm}^2$. This mismatch results in approximately 0.36 extra counts per \M photon, originating from photons that travel across regions outside the active iron film.

Accounting for all losses---including the branching ratio for 14.4~~keV emission, internal conversion, solid angle, and attenuation in the $500~\mathrm{\upmu m}$ quartz substrate---the resulting net count rate of $14.4$~keV photons is $1.3$~kHz. Combined with background contributions, the total detected rate is $2.2$~kHz, corresponding to an average of $2.2$~Hz per energy bin of the spectrometer. 
\begin{table}
    \centering
    \setlength{\tabcolsep}{12pt}
    \begin{tabular}{l lc}
        \hline
        \\
         Losses & EC branching fraction to 136~keV state & 0.998 \\
         & Probability of the 14.4~keV transition & $0.89$\\
         & Internal conversion survival probability & $\frac{1}{1+8.56}=0.11$\\
         & Solid angle fraction & $3.2\cdot10^{-5}$ \\
         & Attenuation of 14.4~keV $\gamma$s in $500\,\mathrm{\upmu m}$ $ \mathrm{SiO}_2$ (device substrate) & 0.42\\
         &  & ------------------------\\
         & Detected fraction of 14.4keV-$\gamma$ per $^{57}$Co decay  & $1.3\cdot10^{-6}$\\
         \\
         \hline
         \\
         Rates & $^{57}$Co source activity   &\hspace{-3cm}1~GHz \\
         & Rate of detected 14.4keV & 1.2~kHz\\
         & Rate of Compton events & 0.5~kHz\\
         & Collimator-Fe overlap mismatch & 0.4~kHz\\
         &  & ------------------------\\
         & Total detected rate & 2.1~kHz\\
         & \\
    \end{tabular}
    \caption{Estimate of losses in the spectrometer and rate of detected 14.4~keV $\gamma$s.}
    \label{tab:my_label}
\end{table}

\FloatBarrier
\subsection{Baseline and Peak fitting}\label{sec:fitting}

\begin{figure}[h]
    \centering
    \includegraphics[width=18cm]{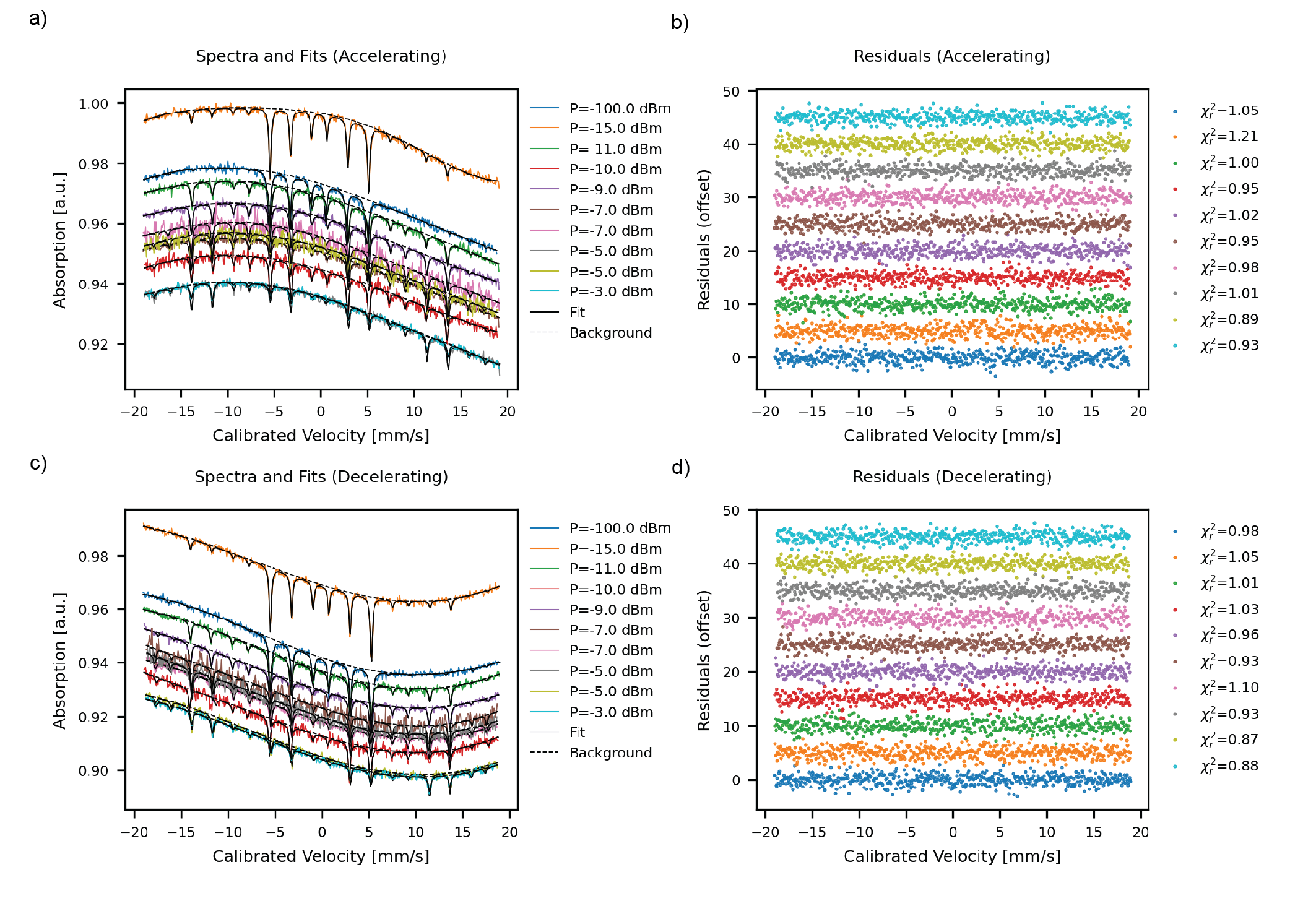}
    \caption{Measured spectra for all modulation settings while the source is accelerating and (a) and decelerating (c). Black lines are the fits to the curves, while the dashed lines represent the fit to the baseline only. b) and d) show the normalized residuals (each offset by 5 for clarity) and reduced chi-square $\chi^2_\nu$ value for each fit.}
    \label{fig:figureSI_5}
\end{figure}

To accurately extract spectral features in the presence of non-ideal detector response and source noise, each acquired \M spectrum is fitted with a model composed of a polynomial background and a sum of the expected 18 Lorentzian peaks. This approach allows to decouple the slowly varying baseline from the narrow resonant absorption features associated with the hyperfine transitions of $^{57}$Fe.
The baseline is modeled with a fifth-order polynomial, capturing slow variations due to inhomogeneous illumination or detector drifts. Each resonance peak is modeled as a Lorentzian. The linewidth is kept fixed across all peaks, since linewidth variations across the peaks are subdominant. In the absence of SAW modulation, the model includes six Lorentzian lines corresponding to the allowed magnetic dipole transitions between hyperfine-split nuclear levels. Under SAW drive, each parent line is replicated into multiple sidebands, symmetrically spaced by integer multiples of the SAW frequency. The number and intensity of visible sidebands depend on the modulation index and follow a squared Bessel distribution, as detailed in Eq. (1) of the main text. For symmetry reasons (see Figure~\ref{fig:figureSI_5}a) the fitting parameters are reduced to 7 Lorentzian amplitudes,  18 Lorentzian resonances, 1 global linewidth, 5 polynomial coefficients for baseline modeling, and 1 offset. 
As shown in Figure~\ref{fig:figureSI_5} the quality of each fit is assessed using the reduced chi-squared statistic, which consistently yields values close to unity, indicating that the model captures the essential spectral features within statistical uncertainties. Residuals between the data and fit show no systematic structure and are randomly distributed across the velocity range, further supporting the validity of the fitting model and confirming that instrumental and background effects are well accounted for. Note that the accelerating and decelerating part of the cycle are calibrated separately and each side contributes as an independent measurement.

\subsection{Velocity Calibration}

\begin{figure}[h]
    \centering
    \includegraphics[width=18cm]{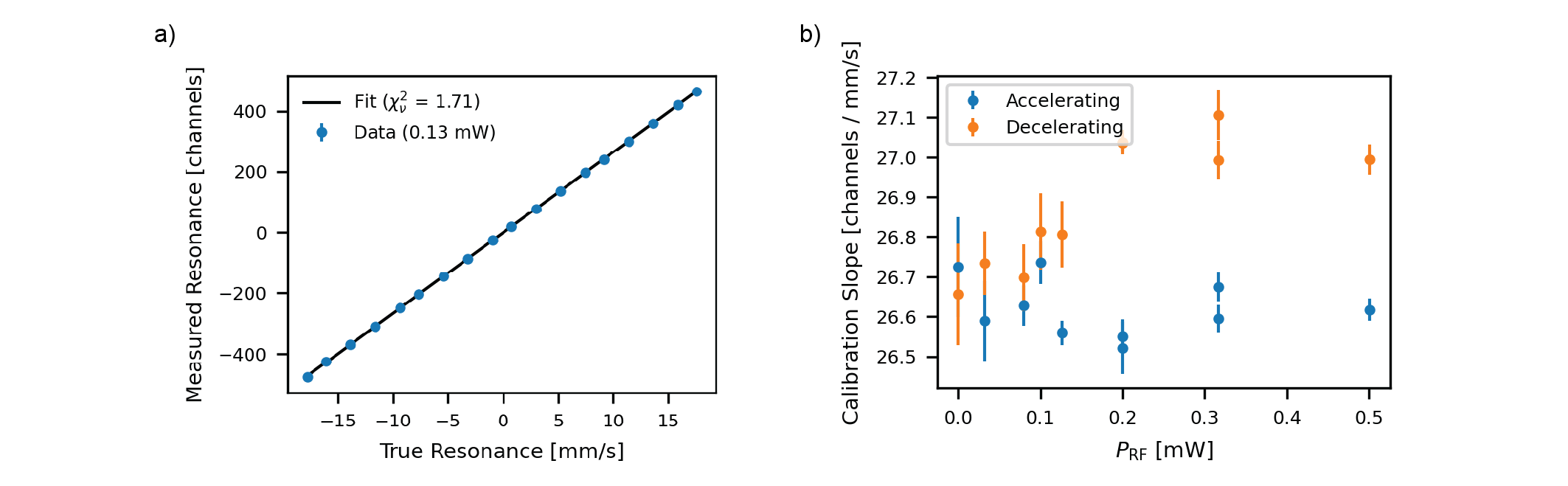}
    \caption{a) Example of velocity calibration curve. b) Calibration slope measure for each data point. The variation is probably determined by environmental conditions.}
    \label{fig:figureSI_6}
\end{figure}

To determine the relationship between channel number and effective Doppler velocity, the spectrometer is calibrated using the known positions of the six Zeeman-split hyperfine transitions in $\alpha$-Fe. These transitions are spaced by 24.4 MHz (101 neV), corresponding to Doppler shifts of $\pm2.1$~mm/s.
The expected velocities for the six resonances in $\alpha$Fe, relative to a $^{57}$Co source embedded in a Rhodium  matrix, are : $\{-5.42, -3.19, -0.95, +0.72, +2.96, +5.19\}$~mm/s~\cite{violet_mossbauer_1971,longworth_preparation_1971}. This spectral pattern is expected to repeat at offsets of $\pm \,n\cdot 8.244$~mm/s, corresponding to the SAW-induced sidebands of order $n$. These values define the reference positions against which we compare the measured absorption peaks recoded from fitting the spectrum  in Section~\ref{sec:fitting}.
The calibration yields a velocity scale that is linear across the full scan range ($\pm19$~mm/s) as shown in Figure~\ref{fig:figureSI_6} a). Figure~\ref{fig:figureSI_6} b) shows that the velocity calibration factor of our spectrometer varies by less than 2\% over the whole dataset (10 days per point), likely due to environmental conditions.

\FloatBarrier
\section{Data Analysis}
\subsection{Sideband Power Estimation}\label{sec:sideband}
\begin{figure}[h]
    \centering
    \includegraphics[width=18cm]{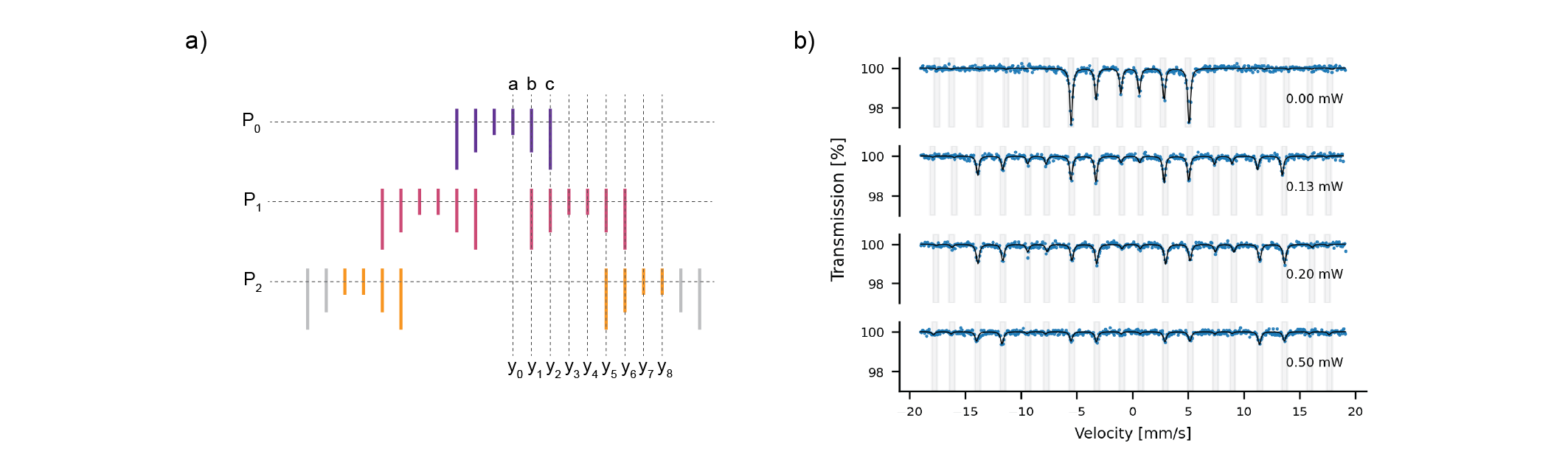}
    \caption{a) Schematic representation of the contribution of the sidebands $P_i$ of each fundamental absorption peak $\{a,b,c\}$ to the 16 (8 for symmetry) distinct peaks $y_j$. The grayed out lines fall out of the measurement range b) Four examples of the measured spectra, normalized by the fitted baseline, and the integration area considered for each peak around the resonances extracted from the fit.}
    \label{fig:figureSI_7}
\end{figure}

We estimate the sideband power by analyzing the spectral weight redistributed by the acoustic modulation of the absorber.
Once the locations of the 18 peaks in the modulated absorption spectrum have been identified, the power in each sideband is extracted by integrating the area of these peaks over a bandwidth of twice the linewidth $\Gamma$. The resulting values are labeled symmetrically as $\mathbf{y} = [y_0, y_1, \dots, y_8]$. This integration approach provides results that are comparable to those obtained by fitting the peak contrast, but it is less sensitive to systematic uncertainties such as drifts in the velocity calibration, which can broaden peaks and reduce their effective contrast.
Each of these peaks originates from the convolution of the six Zeeman-split nuclear transitions with the acoustic sidebands generated by the surface acoustic wave (SAW). By measuring the relative intensities of the unmodulated nuclear transitions as $a$, $b$, and $c$, the matrix $\mathbf{A}$ mapping the unknown sideband powers $\mathbf{x} = [ P_0, P_1, P_2 ]$ to the observed peak integrals is constructed. This relationship is expressed as (see Figure~\ref{fig:figureSI_7}a):
\begin{equation}
\mathbf{y} = \mathbf{A} \cdot \mathbf{x}
\label{eq:matrixproduct}
\end{equation}
with the matrix $\mathbf{A}$ defined by:
\begin{equation}
\mathbf{A} = \begin{bmatrix}
a & 0 & 0 \\
b & c & 0 \\
c & b & 0 \\
0 & a & 0 \\
0 & a & 0 \\
0 & b & c \\
0 & c & b \\
0 & 0 & a \\
0 & 0 & a
\end{bmatrix}
\end{equation}
The overlap between peaks from different sidebands is a consequence of the specific choice of SAW drive frequency and the resolution of our spectrometer.
By inverting Equation~\ref{eq:matrixproduct}, the sideband powers for each measurement is calculated:
\begin{equation}
\mathbf{x} = \mathbf{M} \mathbf{y}, \quad \text{where} \quad \mathbf{M} = \mathbf{A}^{-1} \quad \text{or} \quad \mathbf{M} = (\mathbf{A}^T \mathbf{A})^{-1} \mathbf{A}^T
\end{equation}
The uncertainties in the estimated sideband powers are propagated from the measurement uncertainties encoded in the covariance matrix $\Sigma_y = \mathrm{diag}(\sigma_y^2)$, using standard linear error propagation:
\begin{equation}
\Sigma_p = \mathbf{M} \, \Sigma_y \, \mathbf{M}^T
\end{equation}
This yields the covariance matrix $\Sigma_p$ for the inferred sideband areas $\mathbf{p}$.

\subsection{Global Fit of Sideband Amplitudes}

To quantitatively interpret the evolution of the sideband amplitudes with increasing SAW drive power, a global fit of the three measured datasets corresponding to sidebands of order 0, 1, and 2 was performed. The model assumes that the local nuclear transition experiences a sinusoidal Doppler modulation due to the SAW, producing sidebands whose relative amplitudes are given by squared Bessel functions $J_n^2$. However, because the SAW amplitude varies spatially due to partial reflections and standing-wave formation, the total response is modeled by integrating the Bessel contributions over an inhomogeneous amplitude distribution.
Specifically, the local SAW amplitude is assumed to follow a square-root-transformed cosine profile, leading to an asymmetric distribution $p(y)$ for the amplitude $y$ (see Section \ref{sec:theory}). In practice, this results into an inhomogeneous amplitude profile, the power in each sideband $P_n(x)$ can be numerically integrated. For each sideband order $n$ a
nd RF drive amplitude $x = \sqrt{P_{in}}$, the predicted signal is given by:
\begin{equation}
    P_n(x) = \int  p(y)\, J_n^2(m x y) \,dy.
\end{equation}

\begin{figure}[h]
    \centering
    \includegraphics[width=18cm]{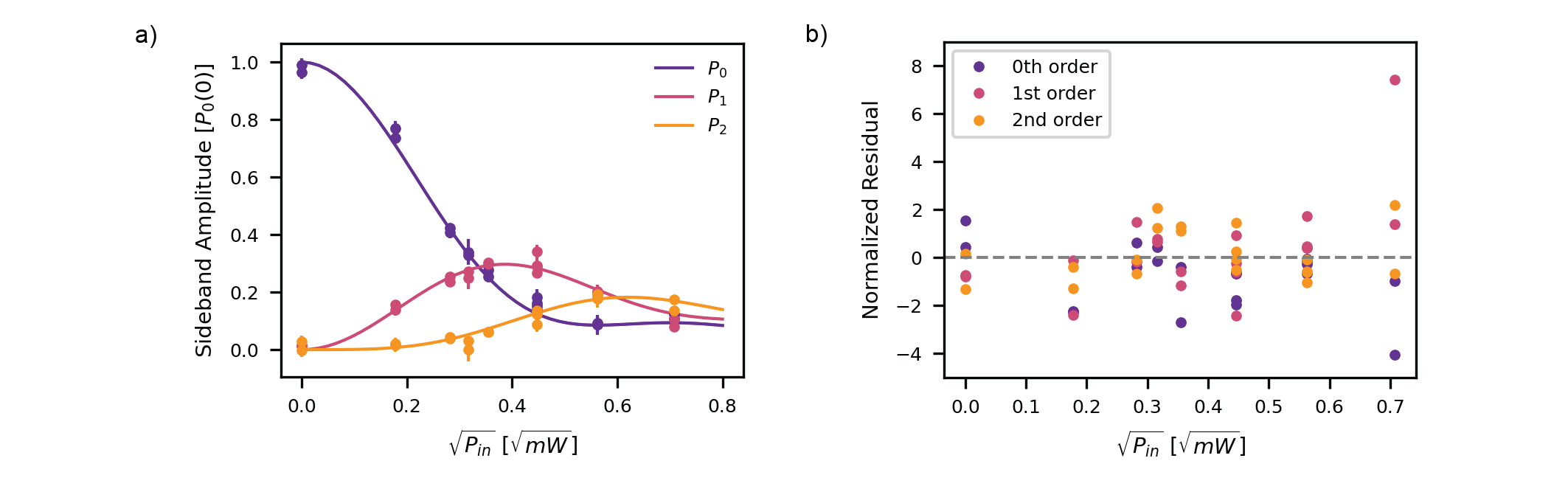}
    \caption{a) Reconstructed amplitudes of the first three sidebands accessible within the spectrometer range, as a function of the applied SAW drive power. The solid lines are a global fit to all three datasets using a model based on the squared Bessel functions integrated over the SAW amplitude distribution. The two fit parameters are the out-of-plane displacement constant $C_\perp$ and the reflection coefficient $\alpha$, which encodes the degree of standing-wave formation due to SAW reflections. b) Normalized residuals for the global fit. The two outlying points responsible for the large $\chi^2_\nu$ are visible at the highest power for the $0^\mathrm{th}$ and $1^\mathrm{st}$ order sidebands.}
    \label{fig:figureSI_8}
\end{figure}

 The parameters $m$ (a global scaling factor between RF amplitude and mechanical modulation) and $\alpha$ (describing the degree of SAW inhomogeneity) are fit globally across all powers and sideband orders. The fit minimizes the residuals between model predictions and experimental data, normalized by the measurement uncertainties.
The best-fit values obtained from least-squares optimization are:
\begin{equation}
m = 4.34 \pm 0.30, \qquad \alpha = 0.36 \pm 0.02,
\end{equation}
Where the error is dominated by a $6\%$ systematic uncertainty drive amplitude. The reduced chi-squared of the global fit is $\chi^2_\nu=2.56$, indicating moderate tension between the model and the data. However, this value is dominated by just two outlying data points(see Figure~\ref{fig:figureSI_8}b); removing these brings the reduced chi-squared down to $\chi^2_\nu=1.38$, consistent with statistical noise and within experimental uncertainty. The value of $\alpha$ from the fit is in excellent agreement with the one ($0.34$) calculated in section~\ref{sec:s-parameters}.
From the fitted value of $m$, the out-of-plane displacement constant $C_\perp$ is extracted. This sets the proportionality between applied RF power and SAW amplitude via the relation $A = C_\perp\, \eta \sqrt{P}$. Using the known values of the SAW wavevector $k_0 = 7.3 \times 10^{10}~\mathrm{m}^{-1}$ and attenuation coefficient $\eta = 0.35$:
\begin{equation}
C_\perp = (5.35\pm 0.37) \cdot 10^{-9}~\mathrm{m/\sqrt{W}}.
\end{equation}
This value quantifies the device’s mechanical response to RF excitation and provides a consistent basis for modeling the evolution of the \M sideband spectrum across the full power range.

\end{document}